\documentstyle[12pt]{article}

\begin{document}

\begin{titlepage}

\begin{flushright}
gr-qc/0008010 \\
$~$ \\
March 2000
\end{flushright}   

\begin{centering}
\vspace{.1in}

\baselineskip=18pt

{\large {\bf PLANCK-LENGTH PHENOMENOLOGY}\footnote{{\bf This paper
received an ``honorable mention'' in the Annual 
Competition of the Gravity Research Foundation for the year 2000.}}}

\vspace{.5in}
{\bf 
Giovanni AMELINO-CAMELIA} \\

\vspace{.2in}

{\small
Dipartimento di Fisica,
Universit\'{a} di Roma ``La Sapienza'',
P.le Moro 2, I-00185 Roma, Italy}

\vspace{.3in}
\vspace{.1in}
{\bf Abstract}\\
\end{centering}

\vspace{.05in}

\baselineskip=16pt

{\small
This author's recent proposal 
of interferometric tests of Planck-scale-related
properties of space-time is here revisited
from a strictly phenomenological viewpoint.
The results announced previously are rederived using
elementary dimensional considerations.
The dimensional analysis is then extended to the other two
classes of experiments (observations of neutral kaons
at particle accelerators and
observations of the gamma rays we detect from distant
astrophysical sources) which have been recently considered
as opportunities to explore ``foamy" properties
of space-time.
The emerging picture suggests that there is an objective
and intuitive way to connect the sensitivities of these
three experiments with the Planck length.
While in previous studies the emphasis was always on
some quantum-gravity scenario and the analysis was
always primarily aimed at showing that the chosen scenario would 
leave a trace in a certain class of doable experiments, 
the analysis here reported takes as starting
point the experiments and, by relating in a direct quantitative
way the sensitivities to the Planck length,
provides a model-independent description of the status
of Planck-length phenomenology.}

\end{titlepage}

\newpage

\baselineskip 12pt plus .5pt minus .5pt

\pagenumbering{arabic}
\setcounter{page}{1}
\pagestyle{plain} 


The realization that the magnitude of quantum-gravity effects
is likely to be suppressed by some power
of the minute Planck length $L_p$ ($\sim 10^{-35}m$)
has traditionally led to the expectation
(see, {\it e.g.}, Ref.~\cite{nodata}) that it
might be impossible to 
obtain pertinent experimental information.
Some recent
studies~\cite{ehns,grbgac,gacgwi}
(also see follow-up work
in Refs.~\cite{neutralk,grbpaps,gwipaps})
have however suggested that certain experiments
are finally reaching a level of sensitivity such that
some quantum-gravity effects could in fact be seen,
at least if the magnitude of these effects is estimated
using the most ``optimistic" (here optimism equals
large estimated effects)
theoretical quantum-gravity scenarios.
These previous studies were focused on
one or another quantum-gravity scenario and the analysis was
primarily aimed at showing that the chosen scenario would 
leave a trace in a certain class of doable experiments.
This note examines the proposals~\cite{ehns,grbgac,gacgwi}
(focusing primarily on this author's recent~\cite{gacgwi})
with the objective of extracting
a model-independent characterization
of the experimental sensitivities
that are being reached.
It will be shown that these sensitivities
can be expressed very intuitively
in terms of the Planck length, in ways that appear to justify
a description of 
the present phase of exploration of space-time properties as
the first steps of ``Planck-length
phenomenology".\footnote{It is 
here proposed to adopt the denomination ``Planck-length
phenomenology", instead of the previously used~\cite{polonpap}
``quantum-gravity phenomenology", as a way to
emphasize that the corresponding area of research is motivated
by the fact that some experiments recently managed to
achieve Planck-length-related sensitivities, more than
by any detailed conjecture on the short-distance structure of 
space-time.
While in this note 
the focus is on experiments relevant for space-time-foam
studies, Planck-length phenomenology
is also active on other aspects of the interplay between
gravitation and quantum mechanics 
(see, {\it e.g.}, Refs.~\cite{stringcosmo,ggrev,danew,polonpap}
and references therein).}

The interferometric studies proposed in Ref.~\cite{gacgwi}
were motivated by the observation that
the sensitivity of modern interferometers, even
the 40-meter interferometer
already in operation at Caltech~\cite{ligoprototype},
is such that one can already rule out~\cite{gacgwi,polonpap}
the possibility of
fluctuations in the length $L$ of the arms of these interferometers
that are of Planck-length magnitude and occur at a rate of
one per Planck time.
Of course, Planck-length-related sensitivity to distance fluctuations
can be potentially significant for quantum gravity.
The interplay between gravitation and quantum mechanics
could plausibly result in a picture of space-time which at short
distances is somewhat fuzzy and indeed
involves distance fluctuations.
This is for example true of most quantum-gravity
approaches hosting some form
of ``space-time foam"~\cite{wheely,hawkfoam}.
The observation reported in Refs.~\cite{gacgwi,polonpap}
indicates that
the sensitivity of interferometers is finally reaching a point
where it can be related rather naturally to the Planck length;
however, the analysis that allows a comparison between the data
and the mentioned random-walk scheme of distance fluctuations
is quite involved and some of the intuition for the significance
of these experimental achievements can be lost through the
lines of the derivation.
With some algebra one shows~\cite{gacgwi,polonpap}
that random-walk
fluctuations of Planck length magnitude occurring at a rate of
one per Planck time
correspond to a ``strain noise power
spectrum"\footnote{As discussed rather pedagogically in
Ref.~\cite{saulson},
the power spectrum $\rho_h(f)$ of the strain noise
that would be induced in the length of
the arms of an interferometer
is the most convenient way to characterize models
of distance fluctuations. 
In fact, the strain $h \equiv \Delta L/L$ 
is a measure of the displacement $\Delta L$ in a given
distance $L$, and the associated strain noise power spectrum
is a spectral function that depends only on the frequency
of observation and contains the most significant information
on the nature of the distance fluctuations (such as the mean
square deviation, which can be obtained as the integral of
the power spectrum over the bandwidth
of observation~\cite{saulson}).}
$\rho_h = c L_p L^{-2} f^{-2}$,
and then one observes~\cite{gacgwi}
that for $L \sim 40m$ and $f \sim 500 H\!z$
this formula leads to a prediction that is inconsistent
with the noise-level measurements~\cite{ligoprototype}
performed by the Caltech 40-meter.

An exercise that might contribute to the development of a more
intuitive understanding of these results is the one 
of realizing that a simple dimensional analysis,
without any detailed computation, suffices to show that
a formula of the type $\rho_h = c L_p \Lambda^{-2} f^{-2}$,
with $\Lambda$ a length scale to be determined,
must hold for any model with distance fluctuations
that are of random-walk type and induce effects that
are linear in the Planck length.
This follows from taking into account that
$\rho_h$ carries dimensions of $H\!z^{-1}$
and that there is a general relation~\cite{rwold} between
stochastic processes of random-walk type and $f^{-2}$
behaviour of the power spectrum.
It is therefore proper to state that the theoretical aspects
of the analysis reported in Refs.~\cite{gacgwi,polonpap}
are only necessary in order to fix the value of $\Lambda$:
they show that $\Lambda = L$ for a random-walk scheme
with Planck-length fluctuations in the length $L$
of the arms of the interferometer occurring at a rate of
one per Planck time.
The phenomenological aspects of the analysis reported in
Refs.~\cite{gacgwi,polonpap} show that the data
reported in Ref.~\cite{ligoprototype}
imply that the prediction $\rho_h = c L_p \Lambda^{-2} f^{-2}$
is inconsistent with available experimental data
unless $\Lambda \gg L$ (which is a rather significant bound
since $L$ is the largest
physical length scale in the experimental context).

This more intuitive reinterpretation of the results
reported by this author in Refs.~\cite{gacgwi,polonpap}
can be cast into an even more general perspective by
noticing that there is a simple sense in which the sensitivity
of interferometers is very clearly at the level of effects
linear in the Planck length. In fact, modern interferometers
are rapidly approaching sensitivity levels corresponding to
strain noise power spectrum of order
$\rho_h \sim 10^{-44} H\!z^{-1}$ (the sensitivity 
which the LIGO/VIRGO generation~\cite{ligo,virgo} of interferometers
is planning to achieve, at least in a relatively narrow
bandwidth, already in their ``first phase
of operation") and this sensitivity level is quite naturally
expressed as the ratio between the Planck length
and the speed-of-light
constant $L_p/c \sim 5 {\cdot} 10^{-44} H\!z^{-1}$
({\it i.e.} the ``Planck time").

It is somewhat
amusing to notice that these interferometers have been
motivated to reach sensitivities in the neighborhood
of $10^{-44} H\!z^{-1}$
because their primary objective
is the discovery of the classical-physics
phenomenon of gravity waves, predicted by
Einstein's general relativity,
and it just happens to be the case that
the relevant classical-physics studies
have led to the conclusion that
a sensitivity somewhere between $10^{-44} H\!z^{-1}$ 
and  $10^{-46} H\!z^{-1}$
is needed for the discovery of classical gravity waves.
It is a remarkable numerical accident that the result of these
classical-physics studies (in which, in particular, the
magnitude of the effects has a strong dependence on the expected
distance between our detectors
and the relevant astrophysical sources of gravity waves)
pointed us toward a sensitivity level
which I am now observing to be also naturally associated with the
intrinsically quantum scale $L_p/c$.

One other reason of interest in the use of interferometers
in space-time foam studies is that quantum-gravity noise 
would be a property of the apparatus.
This might be the first chance to explore whether indeed there
are drastically new phenomena~\cite{polonpap} 
associated with the quantum properties
of the gravitational degrees of freedom of an apparatus.
Interferometers are so sensitive that already the
quantum properties of non-gravitational degrees of freedom
of the apparatus play a non-trivial role~\cite{saulson},
and it appears legitimate to wonder whether
quantum properties
of the gravitational degrees of freedom of the apparatus
are also significant. 
Such properties, when described operatively as in Ref.~\cite{polonpap},
are the first operatively-defined
properties of space-time foam (which is otherwise an intuitive
but abstract concept).

Considering now
the other two quantum-gravity experiments,
the study of foam-deformed neutral-kaon
dynamics using particle accelerators~\cite{ehns,neutralk}
and the study of foam-deformed photon propagation using data on
the gamma rays we detect from distant astrophysical
sources~\cite{grbgac,grbpaps},
it is important to observe that
an intuitive understanding based on dimensional
analysis in terms of fundamental constants
is also possible.
Tests of CPT symmetry in the neutral kaon system
are now~\cite{ehns,neutralk}, at the sensitivity
level $|M_{{K}_0} - M_{{\bar K}_0}|/M_{{K}_0} < 10^{-18}$.
Even without a specific quantum-gravity model it is
easy to realize that it would not be surprising to find some
sort of CPT-violating effects in quantum gravity (which
in particular might involve some form of non-locality,
quantum decoherence, and other ``CPT-theorem-hostile" features)
and a straightforward dimensional
analysis allows to estimate that the magnitude of
such effects in the neutral kaon system, if linear
in $L_p$, would be naturally set 
by $L_p c M_{{K}_0}/ \hbar \sim 5 {\cdot} 10^{-19}$.
Also in this case of neutral-kaon studies
of CPT-violating effects it is therefore proper to state that
the sensitivities being reached correspond to
the natural strength of effects which could emerge
within the hypothesis of linearity in the Planck length.

Tests of Poincar\'{e} symmetries using data on the gamma rays
we detect from distant astrophysical sources 
are rapidly becoming more and more sensitive
with the improvement of
the techniques of signal analysis and source-distance measurement.
In particular, the limits on the possibility of an
energy-dependence of the time of arrival of photons of different
energies emitted simultaneously by a distant astrophysical source
are now~\cite{grbgac,grbpaps}, depending on the particular
phenomenon being studied, somewhere between the
levels $\Delta T/T \sim 10^{-22}$, which is within
the reach~\cite{grbgac} of gamma-ray-burst experiments
analyzing photons with energies
typically in the range $100KeV < E < 1MeV$,
and $\Delta T/T \sim 10^{-16}$,
which is within
the reach~\cite{grbgac,grbpaps} of experiments
analyzing the very-high-energy (multi-TeV) photons
we receive from other types of astrophysical sources.
(Here the intuitive notations $T$ and $\Delta T$
have been adopted respectively for the overall time of travel
and the difference in the times of arrival of two photons
within the signal.)
Even without a specific quantum-gravity model it is
easy to realize that it would not be surprising to find some
form of Poincar\'{e}-violating effects in quantum gravity
(as discussed rather 
pedagogically in Ref.~\cite{polonpap,thooft,gacgrf98,gampul},
these are indeed quite natural because of the element of
discreteness introduced by the Planck length);
these in turn would affect the structure of the dispersion
relations which characterize the laws of particle propagation,
quite possibly leading to an energy-dependence of these
laws~\cite{grbgac,thooft,gampul}.
Working again in the hypothesis of effects linear in the Planck
length, and taking into account the fact that this particular
effect should manifest a dependence on the energy
difference $\Delta E$ between the
two photons whose travel times are being compared,
it is easy to identify the dimensionless ratio that could
set the magnitude of this quantum-gravity effect:
$\Delta T/T \sim L_p \Delta E/(c \hbar)$.
For $\Delta E \sim 1 MeV$,
as found in some gamma-ray-burst data,
this would lead to the estimate $\Delta T/T \sim 10^{-22}$
which is just of the order of the ``$\Delta T/T$ sensitivity"
being reached~\cite{grbgac} by gamma-ray-burst experiments.

The model-independent characterization
of the sensitivities of the
three mentioned ``quantum-gravity experiments'',
which is the main objective of the present note,
is now complete.
It is worth emphasizing again that
the three experiments all happen to have sensitivities
corresponding to the estimates one obtains assuming
that the magnitudes of the new effects depend linearly
on the Planck length.
This coincidence is even more remarkable
in light of the fact that not long
ago the sensitivities of the three
experiments were several orders of magnitude
away from the present levels
and they were all different from one another.
The fact that these experimental sensitivities have improved
so much and now happen to cross together the significant
milestone represented by effects linearly suppressed by the
Planck length gives added significance to the present phase
of Planck-length phenomenology.

One may wonder what makes these three experiments so special
that they managed to reach such an
extraordinary level of sensitivity.
This is easily understood as being mostly related to the
fact that the three experiments are all structured in ways that
render them capable of detecting some ``collective
effects", effects which are the end result
of a large number of minute quantum-gravity effects.
The gloomy forecasts for Planck-length phenomenology
which one finds in traditional quantum-gravity reviews
are based on the observation that under ordinary conditions
the direct detection of a single quantum-gravity effect
would be well beyond our capabilities,
if the magnitude of the effect is suppressed by
the smallness of the Planck length ({\it e.g.},
in particle-physics contexts
the graviton contribution 
to ordinary scattering processes
is completely negligible).
However, small effects can become observable when the 
experimental setup is such that
a very large number of the
small effects is somehow put together.
This later possibility is not unknown to the particle-physics
community, since it has been
exploited~\cite{gacproton} in the context of
proton-decay experiments,
where stringent bounds are achieved by monitoring a very
large number of protons.
An interferometer with bandwidth roughly centered at $100H\!z$
is collecting high-quality data 
over a time scale of order $10^{-2}s$,
and can therefore in principle be sensitive to
the collective effect of an extremely large number of distance
fluctuations if these fluctuations occur with a characteristic
rate of one per Planck time ($\sim 5 {\cdot} 10^{-44}s$).
For example, in the mentioned random-walk $L_p$-linear
scheme of distance fluctuations the length scale
that most genuinely characterizes the associated noise
at $100H\!z$ is $\sqrt{L_p 10^{-2}s} \sim 10^{-14}m \gg L_p$.
Similarly, photons reaching us from distant astrophysical
sources have traveled for a very long time
and can in principle be sensitive to the collective
effect of an extremely large number of ``interactions"
with the space-time foam, if these are relatively frequent.
The case of studies of neutral kaons is even more interesting.
The effect turns out to be testable because of two ``positive
factors": the kaon lifetime, while being significantly shorter
than astrophysical propagation times,
is still significantly larger than the
Planck time (thereby providing an opportunity for sensitivity
to the collective effect of a large number of interactions
with the space-time foam),
and, in addition, the neutral-kaon system hosts a very delicate
balance of mass scales, reflected in peculiarly small dimensionless
ratios such
as $|M_{{K}_L} - M_{{\bar K}_S}|/M_{{K}_0} \sim 10^{-14}$,
which renders it very sensitive~\cite{ehns,neutralk}
to certain types of new phenomena.

Even though the analysis reported in the present note is strictly
phenomenological, this author hopes to have provided a
tool that can be useful
also for work on the quantum-gravity formalism:
while previous phenomenological analyses
might have left those working on the formalism uncertain
on whether or not these developments could be relevant for
their chosen approach, the present model-independent
characterization should give a more intuitive picture
of the sensitivities. This might in some cases be sufficient
for the identification of the elements
of the formalism that are needed in order to make the
connection with phenomenology. ``Canonical/Loop Quantum 
Gravity"~\cite{canoloop} and Superstring Theory
might be already mature enough for this type of analyses.
A key problem remains the one of developing
a plausible measurement theory
for these quantum-gravity approaches.
In ``Canonical/Loop Quantum 
Gravity" some important open issues\footnote{One important
missing element is a compelling candidate as
length observable.
This has wide implications for all
physical predictions, since most measurement procedures
require some distance measurements (as illustrated
by the example considered in Ref.~\cite{areapap},
some distance measurements naturally intervene at intermediate
stages of the procedure,
even when the observable in which
one is finally interested is not a distance).}
with respect to measurement theory
have already been discussed in Ref.~\cite{areapap}.

Going back to the phenomenological viewpoint for the
closing remarks, it is perhaps worth emphasizing that
the chances of discovery of new physics
by Planck-length phenomenology appear to depend very strongly on
whether or not there is in Nature an effect whose magnitude goes
linearly with the Planck length. The three experiments
considered in the present analysis probe space-time in rather
complementary ways but all have sensitivities corresponding to
linearity in the Planck length. It appears therefore likely
that new effects linear in the Planck length would eventually
be discovered,
while instead the case of effects suppressed
quadratically by the Planck length might be beyond our reach.

\baselineskip 12pt plus .5pt minus .5pt


\begin{thebibliography}{99}

\bibitem{nodata} 
C.J. Isham, {\it Structural issues in quantum gravity},
in {\it Proceedings of General relativity and gravitation}
(Florence 1995).

\bibitem{ehns} J.~Ellis, J.S.~Hagelin,
D.V.~Nanopoulos and M.~Srednicki,
{\it Search for violations of quantum mechanics},
Nucl.~Phys.~B241 (1984) 381.

\bibitem{grbgac} G. Amelino-Camelia, J. Ellis, N.E. Mavromatos, 
D.V. Nanopoulos and S. Sarkar, 
{\it Tests of quantum gravity from observations 
of $\gamma$-ray bursts}, 
astro-ph/9712103,
Nature {393} (1998) 763-765.

\bibitem{gacgwi} G.~Amelino-Camelia, 
{\it Gravity-wave interferometers as quantum-gravity detectors},
gr-qc/9808029,
Nature {398} (1999) 216.

\bibitem{neutralk} J. Ellis, J. Lopez, 
N. Mavromatos, D. Nanopoulos and CPLEAR Collaboration,
Phys.~Lett.~B364 (1995) 239;
V.A.~Kostelecky and R.~Potting,
Phys.~Rev.~D51 (1995) 3923;
P.~Huet and M.E.~Peskin,
Nucl.~Phys.~B434 (1995) 3;
F.~Benatti and R.~Floreanini, 
Nucl.~Phys.~B488 (1997) 335.

\bibitem{grbpaps} B.E.~Schaefer,
Phys.~Rev~Lett.~82 (1999) 4964;
S.D. Biller {\it et al},
Phys.~Rev.~Lett.~83 (1999) 2108;
G.~Musser, Scientific American, October 1998 issue.

\bibitem{gwipaps} D.V.~Ahluwalia,
gr-qc/9903074,
Nature 398 (1999) 199;
M.~Brooks,
New Scientist 2191 (1999) 28;
A.~Campbell-Smith, J.~Ellis, N.E.~Mavromatos 
and D.V.~Nanopoulos,
Phys.~Lett.~B466 (1999) 11;
G.~Amelino-Camelia,
gr-qc/9903080, Phys.~Rev.~D62 (2000) 024015;
Y.J.~Ng and H.~van Dam,
gr-qc/9906003;
Hong-wei Yu and L.H.~Ford,
gr-qc/0004063.

\bibitem{polonpap} G. Amelino-Camelia, {\it Are we at the dawn of
quantum-gravity phenomenology?}, gr-qc/9910089, notes based on
lectures given at the XXXV Karpacz Winter School of Theoretical
Physics {\it From Cosmology to Quantum Gravity}, Polanica, Poland,
2-12 February, 1999 (published in the volume entitled ``Towards
Quantum Gravity'', Springer-Verlag Heidelberg 2000, edited by
J.~Kowalski-Glikman).

\bibitem{stringcosmo} R.~Brustein, M.~Gasperini, M.~Giovannini
and G.~Veneziano, Phys.~Lett.~B361 (1995) 45.

\bibitem{ggrev}
G.F.~Giudice, hep-ph/9912279,
{\it Recent developments in physics beyond the standard model},
talk given at the {\it 19th International Symposium on Lepton 
and Photon Interactions at High-Energies}, Stanford, California, 
9-14 Aug 1999 (to appear in the proceedings).

\bibitem{danew} G.Z.~Adunas, E.~Rodriguez-Milla and D.V.~Ahluwalia,
Phys.~Lett.~B485 (2000) 215.

\bibitem{ligoprototype} A.~Abramovici {\it et al},
Phys.~Lett.~{A218} (1996) 157.

\bibitem{wheely} J.A.~Wheeler, {\it Relativity, groups and topology},
ed.~B.S. and C.M.~De Witt (Gordon and Breach, New York, 1963).

\bibitem{hawkfoam} S.W.~Hawking, 
Nuc.~Phys.~B144 (1978) 349.

\bibitem{saulson} P.R.~Saulson, {\it Fundamentals of interferometric 
gravitational wave detectors} (World Scientific 1994).

\bibitem{rwold} V.~Radeka,
IEEE Trans.~Nucl.~Sci.~NS16 (1969) 17;
Ann.~Rev.~Nucl.~Part.~Sci.~38 (1988) 217.

\bibitem{ligo} A.~Abramovici {\it et al}, 
Science {256} (1992) 325. [Updated information
on expected sensitivity of an advanced phase
of the LIGO interferometer can be found
at WWW site http://www.ligo.caltech.edu/~ligo2/.]
 
\bibitem{virgo} C.~Bradaschia {\it et al},
Nucl.~Instrum.~Meth.~{A289} (1990) 518;
B.~Caron {\it et al},
Class.~Quantum Grav.~14 (1997) 1461.

\bibitem{thooft} G.~`t Hooft,
Class. Quant. Grav. {13} (1996) 1023.

\bibitem{gacgrf98} G. Amelino-Camelia, 
Mod.~Phys.~Lett.~A13 (1998) 1319. 

\bibitem{gampul} R.~Gambini and J.~Pullin,
Phys.~Rev.~D59 (1999) 124021.

\bibitem{gacproton} The intuition guiding this author's
contributions to the development of Planck-length phenomenology
originates from an analogy with the strategy being used in
modern proton-decay experiments which was discussed
in G. Amelino-Camelia, 
{\it SO(10) grandunification model with
proton lifetime of the order of $10^{33}$ years}
(Laurea thesis,
Facolt\'{a} di Fisica dell'Universit\'{a} di Napoli, 1990).

\bibitem{canoloop} Recent reviews of
this approach can be found in
A.~Ashtekar,
gr-qc/9901023;
M.~Gaul and C.~Rovelli,
gr-qc/9910079
(notes based  on lectures given 
by C.~Rovelli 
at the XXXV Karpacz Winter School of Theoretical Physics
{\it From Cosmology to Quantum Gravity}, Polanica, Poland, 2-12
February, 1999);
L.~Smolin,
Physics World 12 (1999) 79.

\bibitem{arsarea} A.~Ashtekar, C.~Rovelli, and L.~Smolin, 
Phys.~Rev.~Lett.~69 (1992) 237.

\bibitem{areapap}  G.~Amelino-Camelia,
gr-qc/9804063, Mod.~Phys.~Lett.~A13 (1998) 1155;
gr-qc/9808047,
in {\it Proceedings of 7th International Colloquium 
on Quantum Groups and Integrable Systems}
(Prague, Czech Republic, 18-20 June 1998). 

\end{thebibliography}
\end{document}